\documentclass[12pt,preprint]{aastex}

\usepackage{rotating}

\shorttitle{Free Magnetic Energy}
\shortauthors{R\'egnier and Priest}

\def\nlff{{\em nlff}}
\def\lff{{\em lff}}

\begin{document}

\title{Free Magnetic Energy in Solar Active Regions \\
	above the Minimum-Energy Relaxed State}

\author{S. R\'egnier and E. R. Priest}
\affil{School of Mathematics, University of St Andrews, St Andrews, Fife, KY16
9SS, UK}

\begin{abstract}
	To understand the physics of solar flares, including the local
	reorganisation of the magnetic field and the acceleration of energetic
	particles, we  have first to estimate the free magnetic energy
	available for such phenomena, which can be converted into kinetic and
	thermal energy.  The free magnetic energy is the excess energy of a
	magnetic configuration compared to the minimum-energy state, which is a
	linear force-free field if the magnetic helicity of the configuration
	is conserved. We investigate the values of the free magnetic energy
	estimated from either the excess energy in extrapolated fields or the
	magnetic virial theorem. For four different active regions, we have
	reconstructed the nonlinear force-free field and the linear force-free
	field corresponding to the minimum-energy state. The free magnetic
	energies are then computed. From the energy budget and the observed
	magnetic activity in the active region, we conclude that the free
	energy above the minimum-energy state gives a better estimate and
	more insights into the flare process than the free energy above the
	potential field state.

\end{abstract}


\section{Introduction}

Due to the low value of the plasma $\beta$  (the ratio of gas pressure to
magnetic pressure), the solar corona is magnetically dominated. To describe the
equilibrium structure of the coronal magnetic field when gravity is
negligible, the force-free assumption is then appropriate:
\begin{equation}
\label{eq:fff}
\vec \nabla \wedge \vec B = \alpha \vec B,
\end{equation} 
where $\alpha = 0$ gives the potential (or current-free) field, $\alpha = cst$
gives the linear force-free field (\lff), and $\alpha$ being a function of
space gives the nonlinear force-free field (\nlff). The properties of
force-free fields have been well described \citep[e.g.,][]{wol58, mol69, aly84,
ber85}. \citet{wol58} with general astrophysical configurations in mind derived
two important theorems: (i) in the ideal MHD limit the magnetic helicity is
invariant during the evolution of any closed flux systems, (ii) the minimum
energy state is the linear force-free field conserving the magnetic helicity
\cite[see also][]{aly84, ber85}. \citet{tay86} applied this to laboratory
experiments and hypothesized that in a weak but finite resistive regime the
total magnetic helicity of the flux system is invariant during the relaxation
process to a minimum energy state. According to \cite{wol58}, the relaxed state
is then a linear force-free field. Therefore the free magnetic energy that can
be released during a relaxation process is the excess energy of the magnetic
configuration above the linear force-free field with the same magnetic helicity.

\citet{hey84} were the first to suggest the importance of magnetic helicity and
Taylor relaxation in the solar corona. They extended the Woltjer-Taylor theory
for an isolated structure bounded by magnetic surfaces to one that of a coronal
field in which the field lines enter or leave the volume (through the
photosphere): thus the magnetic helicity is allowed to enter or leave the
corona as the photospheric field changes in time. They also suggested that the
coronal field evolves locally through a set of linear force-free fields with
the field continually relaxing and the footpoint connections continually
changing by small-scale turbulent reconnections, which heat the corona.
Moreover they suggested that, if the magnetic helicity becomes too large, an
eruption takes place in order to expel the excess magnetic helicity. The
coronal heating mechanism by magnetic turbulent relaxation was later developed
into a self-consistent theory \citep{hey92}. Based on a statistical analysis of
vector magnetograms, \cite{nan03} have shown that the relaxation process of
flare-productive active regions is similar to Taylor's theory. Nevertheless,
\cite{reg06} have shown that the magnetic helicity can evolve significantly on
a short time scale (about 15 min) and that the evolution of the coronal magnetic
field is often well described by a series of nonlinear force-free
equilibria. The modelled evolution of global coronal fields by successive
nonlinear force-free equilibria was also investigated by \citet{mac07a, mac07b}.

To better understand the physics of flares, we need to estimate the amount of
magnetic energy available in a magnetic configuration for conversion into
kinetic energy and/or thermal energy in a solar flare. There is no free
magnetic energy in a potential field configuration: this is a minimum-energy
state for a given normal magnetic field at the photosphere, and the magnetic
energy depends only on the distribution and amount of flux through the
photosphere. The linear and nonlinear configurations, however, do have free
energy due to the presence of currents. As shown in \citet{reg07}, the energy
storage in active regions can be (i) in the corona due to the existence of
large-scale twisted flux bundles, or (ii) near the base of the corona
associated with the existence of a complex topology. The free energy can be
estimated from photospheric or chromospheric magnetic fields based on the
magnetic virial theorem \citep{mol69, aly84}, or from reconstructed 3D coronal
fields (often assuming a force-free equilibrium). Using nonlinear force-free
fields, the magnetic energy budget has been estimated before and after a flare
\citep{ble02, reg06}: as expected the authors found that the magnetic energy
usually decreases during the flare. Nevertheless, it strongly depends on the
strength of the flare, on the processes of energy injection (e.g., flux
emergence, flux cancellation, sunspot  rotation) and on the time span between
the reconstructed fields. \citet{ble02} have suggested that Taylor's theory
does not apply to flares and CMEs. The same conclusion has been reached
previously by numerical simulations \citep[see e.g.,][]{ama00}. This can be
understood if (i) the helicity is not conserved during a flare or a CME in the
finite domain of computation due to the injection of helicity through the
photosphere or into the CMEs, (ii) the eruption phenomenon is often localized
in the active region and so does not affect or modify strongly the
nonpotentiality of the field outside the flare surroundings. Note that the
energy flux (or Poynting flux) can be derived from successive magnetic field
measurements when the plasma flows are known \citep[see e.g.,][]{kus02}. The
Poynting flux gives an estimate of the injected energy through the photospheric
surface due to transverse motions and/or flux emergence.

In this letter, we compute the free magnetic energy for different active
regions assuming a nonlinear force-free equilibrium with a reference field
being either the potential field or the linear force-free field, and from the
magnetic virial theorem. We are assuming that at a given time and with the same
boundary conditions the minimum-energy state is given (\citeauthor{wol58}'s
theorem) by the linear force-free field with the same magnetic helicity as the
nonlinear force-free field. We are not here investigating the validity of the
Taylor-Heyvaerts theory in solar active regions.

\section{Selected Active Regions}
\label{sec:obs}
 
In order to compare the different measurements of free magnetic energy, we have
selected four different active regions with different types of activity
(confined flares, flares associated with a CME or filament eruptions) and at a
different stage of their evolution (before or after a flare):

\paragraph{AR8151:} observed on February 11, 1998 at 17:36 UT, this is an old
decaying active region (decreasing magnetic flux and magnetic polarities
diffusing away). A filament eruption associated with an aborted CME was
reported on Feb. 12, but no flare was observed. The vector magnetic field was
recorded by the MEES/IVM \citep{mic96, lab99}. The high values of the current
density imply strongly sheared and twisted flux bundles  \cite[see][]{reg02,
reg04}. Due to the existence of highly twisted flux tubes (with more than 1
turn) and the stability of the reconstructed filament and sigmoid (with less
than 1 turn), the authors concluded that the eruptive phenomena was most likely
to be due to the development of a kink instability in the highly twisted flux
bundles; 

\paragraph{AR8210:} observed on May 1, 1998 from 17:00 to 21:30 UT, this is a
newly emerged active region with a complex topology as described in
\cite{reg06}. A M1.2 flare was recorded on May 1, 1998 at 22:30 UT. The
selected vector magnetogram (MEES/IVM) at 19:40 UT was observed during a 
``quiet'' period between two C-class flares. In \cite{reg06}, the authors 
described the magnetic reconnection processes occurring during this time period
and leading to a local reorganisation of the magnetic field. The reconnection
processes are related to the slow clockwise rotation of the main sunspot or a
fast moving, newly emerged polarity. Following the time evolution during 4
hours, the authors showed that the free magnetic energy decreases during 
the flare over a period of about 15 min, and the total magnetic energy is
slightly increased during this time period; 

\paragraph{AR9077:} this corresponds to the famous Bastille day flare in 2000
\citep{liu01, yan01, fle01, kos01, wan05}. The vector magnetogram was recorded
at 16:33 UT after the X5.7 flare which occurred at 10:30 UT. The active region
was still in the magnetic reorganisation phase after the flare and ``post''-flare
loops were observed in 195\AA~TRACE EUV images. The flare was also associated
with a CME;

\paragraph{AR10486:} this active region is responsible for the main eruptions
observed during the Halloween events (26 Oct. to 4 Nov. 2003). The MEES/IVM
vector magnetogram was recorded on October 27, 2003 at 18:36 UT before the
X17.2 flare which occurred at 11:10 UT on October 28. The flaring activity of
this active region and the associated CMEs have been extensively studied. For
instance, \cite{met05} have shown that the large magnetic energy budget ($\sim
3 ~10^{33}$ erg) on Oct. 29 is enough to power the extreme activity of this
active region. 

For these particular active regions, the reduction of the full Stokes vector to
derive the magnetic field has already been detailed in several articles
\citep[e.g.,][]{reg02, reg06} -- the 180-degree ambiguity in the
transverse component was solved by using the algorithm developed in \cite{can93}
\citep[see also][]{met06}. 

\section{Magnetic Fields}

From the observed vector magnetic field as described in Section \ref{sec:obs},
we extrapolate to obtain three types of coronal magnetic field, each of which has
the vertical component of the magnetic field imposed at the photopshere:

\begin{itemize}

\item[-]{{\em potential field}: there is no current flowing in the magnetic
configuration; this is the minimum-energy state that the magnetic field can
reach when the magnetic helicity is not conserved;}

\item[-]{{\em linear force-free field}: we compute the linear force-free whose
$\alpha$ parameter is chosen so that the total magnetic helicity is the same as
the nonlinear force-free; in other words, this gives the minimum-energy state
that conserves the magnetic helicity;}

\item[-]{{\em nonlinear force-free field}: we use the vector potential
Grad-Rubin-like method \citep{gra58, ama99}. The bottom boundary conditions also
require the knowledge of $\alpha$ in one polarity derived from the transverse
field components: $\alpha = \frac{1}{B_z}~\left( \frac{\partial B_y}{\partial x}
-  \frac{\partial B_x}{\partial y} \right)$.   }

\end{itemize}

In order to have energy values which can be compared, we have imposed the same
closed conditions on the side and top boundaries for each model. To satisfy
these conditions, we surround the vector magnetic field observed by MEES/IVM by
weak field measurements provided by SOHO/MDI line-of-sight observations. The
active region fields are then confined by a surrounding potential field and the
magnetic field decreases from the center of the active region (compatible with
the field vanishing at infinity). The magnetic flux is balanced.

\section{Free Magnetic Energy}

From the 3D coronal magnetic configurations, we can derive the magnetic energy
for the different active regions and different models:
\begin{equation}
E_m = \int_{\Omega}~\frac{B^2}{8\pi}~d\Omega
\end{equation}
in a volume $\Omega$. The free magnetic energy is derived from the nonlinear
({\em nlff}) force-free field using either the potential ({\em pot}) or linear
force-free ({\em lff}) field as reference field:
\begin{equation}
\Delta E_{pot}^{nlff} = E_m^{nlff} - E_m^{pot}
\end{equation}
and
\begin{equation}
\Delta E_{lff}^{nlff} = E_m^{nlff} - E_m^{lff}.
\end{equation}
The \lff~field used here has the same relative magnetic helicity as the
\nlff~field satisfying Woltjer's theorem. That implies that the \nlff~field has
to be computed first, and then the \lff~field is determined by an iterative
scheme to find the $\alpha$ value matching the helicity of the \nlff~field. The
relative magnetic helicity is computed from the \cite{ber84} equation
\citep[see e.g.][]{reg05}:
\begin{equation}
\label{eq:hrel}
\Delta H_m = \int_{\Omega}~(\vec A - \vec A_{pot}) \cdot (\vec B + \vec
B_{pot})~d\Omega
\end{equation}
where $\vec B$ and $\vec A$ (resp. $\vec B_{pot}$ and $\vec A_{pot}$) are the
\nlff~(resp. potential) magnetic field and its associated vector potential
computed in the volume $\Omega$. The relative magnetic helicity given by
Eqn.~(\ref{eq:hrel}) satisfies the closed boundary conditions used by the
Grad-Rubin reconstruction method.

\begin{sidewaystable}
\begin{center}
\caption{Free magnetic energy and relative magnetic helicity for the different
active regions}
\label{tab:nrj}
\begin{tabular}{ccccccccc}
\tableline\tableline
 & $E_m^{nlff}$ & $\Delta E_{pot}^{nlff}$ & $\alpha$ & $\Delta E_{lff}^{nlff}$ & 
 	\multicolumn{2}{c}{$\Delta E_m^{vir}$ (10$^{32}$ erg)} & 
 	$\Delta H_m$ & Comments \\
 & (10$^{32}$ erg) & (10$^{32}$ erg) & (Mm$^{-1}$) & (10$^{32}$ erg) & observed & computed & 
 	(10$^{42}$ Mx$^2$) & \\
\tableline
AR8151 & 0.64 & 0.26 & 0.067 & 0.05 & 1.2 & 0.79 & 0.47  & twisted bundles\\
AR8210 & 10.6 & 0.24 & -0.056 & 0.14 & 1.63 & 0.79 & -4.2 & before C flare\\
AR9077 & 14.2 & 2.21 & -0.015 & 1.62 & 0.48 & 1.25 & -14.6 & post-flare loops \\
AR10486 & 70.5 & 18.05 & 0.021 & 7.23 & 41.7 & 2.62 & 35.1 & before X17 flare \\
\tableline
\end{tabular}
\end{center}
\end{sidewaystable}


For the sake of comparison, we also compute the free magnetic energy derived
from the magnetic virial theorem assuming a force-free field \citep[e.
g.][]{aly89, kli92, met95, met05, whe06}. Considering that the magnetic field
can be decomposed into a potential part and a nonpotential one, $\vec B = \vec
B_{pot} + \vec b$, then following \cite{aly89} the free magnetic energy (above
potential) is:
\begin{equation}
\Delta E_m^{vir} = \frac{1}{4\pi}~\int_{\Sigma}~(xb_x + yb_y)B_z~dxdy
\label{eq:vir}
\end{equation}
in the half-space above the surface $\Sigma$. The free magnetic energy from the
virial theorem only requires the magnetic field distribution on the bottom
boundary. We compute Eqn.~\ref{eq:vir} from either the observed vector magnetic
field (not necessarily force-free) or the reconstructed \nlff~field on the
photosphere. It is important to note that the energy values derived from the
magnetic virial theorem are strongly influenced by the spatial resolution as
mentioned in \cite{kli92}.

Considering that the \lff~field is the minimum energy state of the
\nlff~field, the energy values can be sorted as follows:
\begin{equation}
E_m^{pot} < E_m^{lff} < E_m^{nlff}.
\end{equation}

\begin{figure}
\begin{center}
\plotone{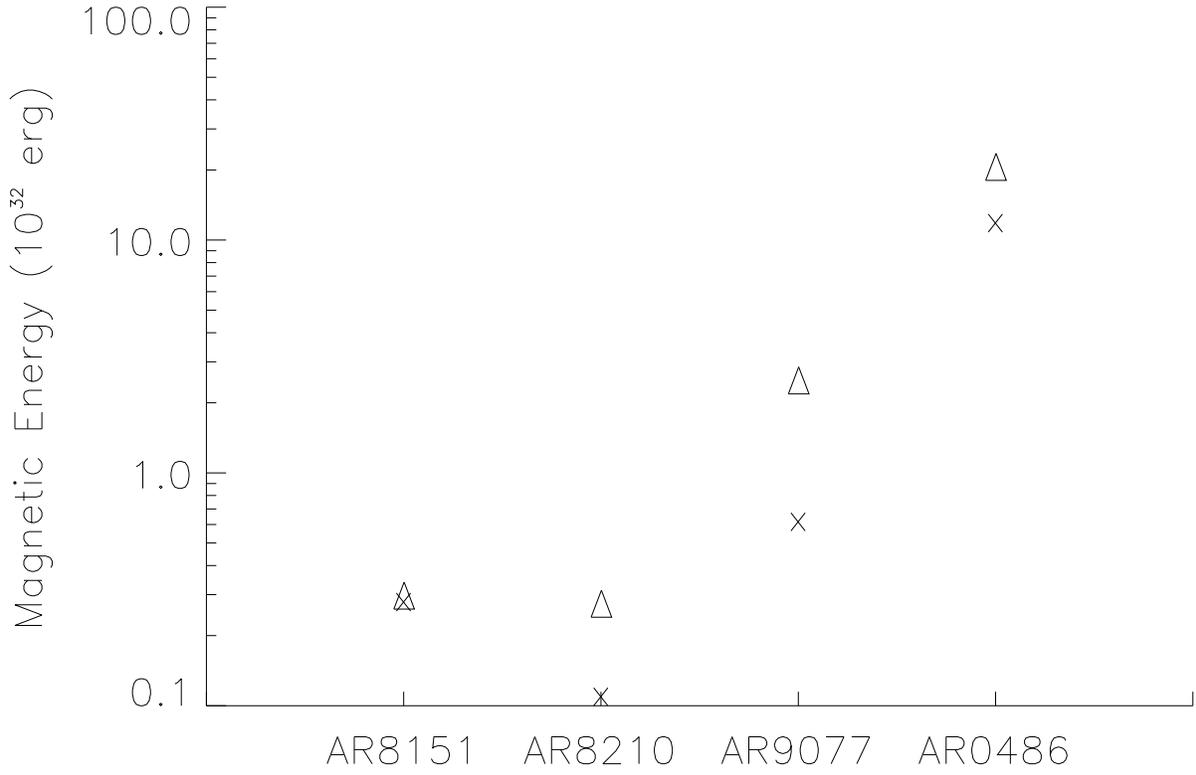}
\caption{Magnetic energy above potential for both the \nlff~field (triangles)
and the \lff~field (crosses) of the four selected active regions (units of
10$^{32}$ erg). The free magnetic energies $\Delta E_{lff}^{nlff}$ are given by
the differences between the triangles and crosses.}
\label{fig:nrj}
\end{center}
\end{figure}

In Fig.~\ref{fig:nrj}, we plot the free energy values in the reconstructed
magnetic configurations using the potential field as reference field for the
\nlff~fields (triangles) and \lff~fields (crosses). The difference between the
two values is the minimum free energy $\Delta E_{lff}^{nlff}$ according to
Woltjer's theorem. Figure~\ref{fig:nrj} clearly shows that the free magnetic
energy can vary by at least 2 orders of magnitude: the energy is strongly
influenced by the total magnetic flux and the distribution of the polarities.
By comparing the amount of free energy $\Delta E_{pot}^{nlff}$ and the observed
eruptive phenomena, we can conclude that $\Delta E_{lff}^{nlff}$ gives a better
estimate of the free energy. For instance, $\Delta E_{pot}^{nlff}$ is similar
for AR8151 and AR8210 but $\Delta E_{lff}^{nlff}$ is nearly three times larger
for AR8210. And the related eruptive phenomena are very different: a slow
filament eruption without a flare for AR8151 and a C-class flare for AR8210.
For AR9077, $\Delta E_{lff}^{nlff}$ is still enough to trigger an X-class flare
but certainly not the X5.7 flare observed prior to the time considered here.
For AR10486, $\Delta E_{lff}^{nlff}$ is significantly reduced compared to
$\Delta E_{pot}^{nlff}$ but still enough to trigger powerful flares which
explains the high level of activity in this active region \citep{met05}.    

In Table~\ref{tab:nrj}, we summarize the different values of free magnetic
energy, the magnetic energy of the \nlff~magnetic configurations ($E_m^{nlff}$)
and the relative magnetic helicity. We also mention the $\alpha$ values used to
compute the \lff~fields satisfying Woltjer's theorem. We notice that the
different values of free energy are consistent and increase when the eruption
phenomena increase in strength with the exception of $\Delta E_m^{vir}$ from
the observed magnetograms. The latter is related to the applicability of the
virial theorem because the observed magnetograms are not force-free at the
photospheric level \citep{met95}.

\section{Conclusions}

We have computed the free magnetic energy from several formulae in various
active regions at different stages of their evolution: from the difference
between  the \nlff~field and either the potential field ($\Delta
E_{pot}^{nlff}$) or  the \lff~field ($\Delta E_{lff}^{nlff}$) having the same
magnetic helicity, and from the magnetic virial theorem ($\Delta E_m^{vir}$)
using either the observed field or the \nlff~field. 

The free magnetic energy $\Delta E_{lff}^{nlff}$ is a better estimate (than
$\Delta E_{pot}^{nlff}$) of the energy budget of an active region available for
flaring assuming that the magnetic helicity is conserved and gives more
insights into the possible eruption mechanisms in the active region. For
AR8151, it is clear that there is not enough energy to trigger a flare capable
of a large-scale reorganisation of the field ($\sim$ 5 10$^{30}$ erg).
Therefore as stated in \cite{reg04} the kink instability of the highly twisted
flux tube is most likely to be responsible for the observed eruptive
phenomenon. Despite a magnetic energy of about 10$^{33}$ erg, the free magnetic
energy in AR8210 is only 1\% of the total energy but is enough to trigger small
confined flares. This is consistent with the observations and modelling
described in \citet{reg06}. We note that for the two possible mechanisms to
store energy \citep{reg07}, the presence of large twisted flux bundles is more
efficient than the highly complex topology: 10\% of free energy in AR8151
compared to 1\% for AR8210. The magnetic energy budget of AR9077 is still
important even if the observed field is after a X5.7 flare. Therefore even
after a strong flare with post-flare loops resembling potential field lines,
the magnetic configuration is far from potential and the energy budget is still
sufficient to trigger further powerful flares.  For AR10486, $\Delta
E_{lff}^{nlff}$ is certainly not sufficient to trigger the observed X17.2
flare, but the $\Delta E_{pot}^{nlff}$ seems to be more consistent with the
recorded flaring activity. This can be explained by the fact that the main
hypothesis of Woltjer's theorem is not satisfied: the X-class flare is
associated with a CME expelling a magnetic cloud (and therefore magnetic
helicity) into the interplanetary medium.  

The free magnetic energy $\Delta E_m^{vir}$ gives  consistent values when
computed from the \nlff~extrapolated fields. For most photospheric
magnetograms, the force-free assumption is not well satisfied and so leads to
inaccurate values of $\Delta E_m^{vir}$ from observations. In \cite{met05}, the
computation of the free energy from the virial theorem was performed using
chromospheric magnetic field measurements which are more force-free than
photospheric magnetograms \citep{met95, moo02}. 

To have a better understanding of flaring activity, our main conclusion is that
it is useful to compute both $\Delta E_{pot}^{nlff}$ and $\Delta
E_{lff}^{nlff}$: the first giving an upper limit on the magnetic energy that
can be released during a large flare, especially when associated with a CME,
the second being a good estimate of the energy budget for small flares and
allowing us to distinguish between different flare scenarios.

\acknowledgments
We thank the UK STFC for financial support (STFC RG). The computations were done
with XTRAPOL code developed by T. Amari (supported by the Ecole Polytechnique,
Palaiseau, France and the CNES). We also acknowledge the financial support by
the European Commission through the SOLAIRE network (MTRN-CT-2006-035484). 


\end{document}